\documentclass[12pt]{iopart}
\usepackage{iopams}
\usepackage{graphicx}
\usepackage{epstopdf}
\def\be{\begin{equation}}
\def\ee{\end{equation}}

\begin{document}

\title{Modified Amplitude of  Gravitational Waves  Spectrum}
\author{Basem Ghayour and P K Suresh}
\address{ School of Physics, University of Hyderabad,
Hyderabad-500 046. India.}
 \eads{\mailto{ph09ph21@uohyd.ernet.in; pkssp@uohyd.ernet.in}}

\begin{abstract}
The spectrum of thermal gravitational waves  is  obtained by including   the high frequency  thermal gravitons  created  from  extra-dimensional  effect and  is   a new  feature  of the spectrum.
The amplitude  and spectral energy density of  gravitational  waves in thermal vacuum state are  found enhanced. 
The amplitude of the waves get modified in the frequency range (10$^{-16}$ -10 $^{8}$ Hz)   but   
the corresponding spectral energy density is  less than  the   upper bound of  various estimated results. 
 With the  addition of   higher frequency thermal  waves,  the  obtained spectral energy density  of the wave in thermal vacuum state does not exceed the upper bound  put by  nucleosynthesis rate.  The existence of  cosmologically originated  thermal gravitational waves  due to extra dimension is not ruled out.
\end{abstract}

\pacs{98.70.Vc,98.80.cq,04.30.-w} 

\maketitle

\section{\label{sec:level1}Introduction }

One of the predictions of general theory of relativity is the existence of gravitational waves. The sources of  generation of these waves  vary  from   the dynamics of early universe to  massive  astrophysical objects  such as  neutron star binaries and black hole mergers etc. Thus the    waves have a  wide spectrum   of frequencies vary from very low to high ${\cal{O}}($10$^{-19}$Hz -10$^{10}$Hz). It is possible to discriminates the  relic waves from other sources on the observational point of view also.   The relic gravitational waves are   paramount importance in cosmology because it provides valuable information on the conditions of very early universe.   It is believed  that the relic gravitational waves are mainly generated during inflationary epoch.  And the waves that amplified  during the inflation are low frequency only.  Since the higher frequency  waves are  outside the  ``barrier'' (horizon)\cite{gh}  \footnote{ the terminology ``barrier" is adopted from \cite{gh}.} the corresponding  amplitudes decreased during the evolution of the universe. 
The features of  relic gravitational  waves  of  very high frequency range  is interesting   though  the energy scale of conventional inflationary models  are not favoring for it. 
However  if  extra dimensions exist (for a review, see \cite{rex} and motivation for extra dimensions \cite{ex}) the graviton background  can have a thermal spectrum \cite{fry}. According to the  extra dimensional models  the  thermal gravitons with very high frequency range also be observed with  a specific peak temperature  today  \cite{fry} and  
 therefore  the detection of very high  frequency thermal gravitational waves is  an interesting test to   see the possibility of existence of extra dimensions as well. 
These  thermal gravitational waves   can contribute to the  higher frequency range of the spectrum.    
The 
existence of  thermal graviton background
with  the black body type spectrum  is   also discussed in \cite{24},\cite{25}. 
If  the inflation was preceded by a radiation era, then there would be  thermal gravitational waves at the time of inflation \cite{25}. The  generation of tensor perturbations during inflation  by  the stimulated emission   process leads  to the existence of  thermal  gravitational waves \cite{29}.  
Direct detection of the thermal gravitons is challenging
but may be possible in the near future with the 21-cm
emission line of atomic hydrogen.

  The inflationary scenario \cite{23} predicts a
stochastic cosmic background of gravitational waves
(CGWB) \cite{24}.
The spectrum of  these relic gravitational waves  depends
not only on the details  of expansion  during the  inflationary era  but also 
 the subsequent stages, including the current  epoch of the universe. 
Computation of  the  spectrum of the  waves   for   matter dominated   universe  is usually  done in  decelerated expanding model \cite{1}-\cite{6}. The resulting spectrum is used  for putting constrain on the detection of  gravitational waves  originated from  sources other than  early universe epoch.
The result of  astronomical
observations on SN Ia \cite{11}-\cite{12} shows that the universe is currently under going  accelerated
expansion  indicating a non-zero cosmological 
constant.  According to the $\Lambda$CDM
concordance model, the observed acceleration of the present universe is supposed to be driven by the dark energy. 
 Effect of  the current acceleration  on the nature  of  the  spectrum  and  spectral energy density of the relic
gravitational waves  is studied  \cite{2},\cite{15}.  And shown that  the current acceleration phase of the universe  does change  the shape, amplitude and spectrum  of the  waves  \cite{15}.

In the present work, we   consider   contribution of  very higher frequency  relic  thermal gravitational waves  to its  spectrum and spectral  energy density  for the decelerated as well as  accelerated universe.  The focus of the present work is on the spectrum of  the higher frequency range of the  waves due to extra dimensional effects.
The  normalization  of the spectrum is being done with the measured  CMB  anisotropy spectrum of the WMAP.   The inclusion of the higher frequency relic thermal gravitational waves    leads to  enhancement of the spectrum.   This  enhancement  leads to  modification of the amplitude of the spectrum  in the frequency range (10$^{-16}$ -10 $^{8}$ Hz)   as an additional  feature  and is  possible to  compare these with the sensitivity of Advanced.LIGO (Adv.LIGO), Einstein Telescope  (ET) and LISA missions.  The corresponding spectral energy density can be compared with estimated upper bound of various studies.
 Also can check whether the  inclusion of  higher frequency thermal gravitational waves   in  the  total spectral energy density  exceed  the  upper  bound   of  primordial nucleosynthesis rate or not.
 In the present work, we use the unit $c=\hbar = k_{B} =1$.

 \section{Gravitational waves spectrum in expanding universe}
The perturbed metric for a homogeneous  isotropic  flat Friedmann-Robertson-Walker (FRW) universe  can be written   as
\begin{equation}
d s^{2}= S^{2}(\eta)(d\eta^{2}-(\delta_{ij}+h_{ij})dx^{i}dx^{j}),
\end{equation}
where $S(\eta)$ is the cosmological scale factor,  $\eta$ is the conformal time and $\delta_{ij}$ is the Kronecker delta  symbol.  The $h_{ij} $ are  metric perturbations  field contain only the pure gravitational waves and is  transverse-traceless i.e; $\nabla_i h^{ij} =0, \delta^{ij} h_{ij}=0$.

The present study mainly deals  with amplitude and spectral energy density of  the  relic gravitational waves   generated by the expanding spacetime  background. Thus  the perturbed matter source is therefore not taken into account. The 
 gravitational waves are described with  the  
 linearized field equation given by
 \begin{equation}\label{weq}
 \nabla_{\mu} \left( \sqrt{-g} \, \nabla^{\mu} h_{ij}(\bf{x}, \eta)\right)=0.
 \end{equation} 
The tensor perturbations have two independent physical degrees of freedom  and are denotes as $h^{+}$ and $h^{\times}$, called polarization modes. To compute the spectrum of gravitational waves $h(\bf{x},\eta)$ in the thermal states, we express $h^{+}$ and $h^{\times}$ in terms
of the creation ($a^{\dagger}$) and annihilation ($a$) operators,
\begin{eqnarray}\label{1}
\nonumber  h_{ij}({\bf x},\eta)=\frac{\sqrt{16\pi} l_{pl}}{S(\eta)} \sum_{\bf{p}} \int\frac{d^{3}k}{(2\pi)^{3/2}} {\epsilon}_{ij} ^{\bf {p}}(\bf {k}) \\
 \times  \frac{1}{\sqrt{2 k}} \Big[a_{\bf{k}}^{\bf {p}}h_{\bf {k}}^{\bf {p}}(\eta) e^{i \bf {k}.\bf {x}} +a^{\dagger}_{\bf {k}} {^{\bf {p}}} h^{*}_{\bf {k}}{^{\bf {p}}} (\eta)e^{-i\bf{k}.\bf{x}}\Big],
\end{eqnarray}
where  $\bf{k}$ is the comoving wave
number, $k=|\bf {k}|$, $l_{pl}= \sqrt{G}$ is the
Planck's length and $\bf{ p}= +, \times$ are polarization modes. The polarization tensor 
$\epsilon_{ij} ^{{\bf p}}({\bf k})$ is symmetric and transverse-traceless  $ k^{i} \epsilon_{ij} ^{{\bf p}}({\bf k})=0, \delta^{ij} \epsilon_{ij} ^{{\bf p}}({\bf k})=0$ and 
satisfy  the conditions $\epsilon^{ij {\bf p}}({\bf k})   \epsilon_{ij}^{{\bf p}^{\prime}}({\bf k})= 2  \delta_{ {\bf p}{{\bf p}}^{\prime}} $ and $ \epsilon^{{\bf p}}_{ij} ({\bf -k}) = \epsilon^{{\bf p}}_{ij} ({\bf k}) $, the creation and annihilation operators  satisfy
$[a_{{\bf k}}^{{\bf p}},a^{\dagger}_{{\bf k} ^{\prime}} {{^{{\bf p}}}^{\prime}}]= \delta_{{{\bf p}} {\bf {p}}^{\prime} }\delta^{3}({\bf k}-{{\bf k}}^{\prime})$, the initial vacuum state is defined  as
\begin{equation}
a_{\bf{k}}^{\bf{p}}|0\rangle = 0,
\end{equation}
for each $\bf {k}$ and $\bf {p}$. The energy density of the gravitational waves in vacuum state is $ t_{00}= \frac{1}{32 \pi l^2_{pl}} h_{ij,0} h^{ij}_{,0}$.

 For a fixed  wave number $\bf{k} $ and a fixed polarization state $\bf{p}$ the linearized wave equation (\ref{weq}) gives
 \begin{equation}\label{zz1}
h^{\prime \prime}_{k}+2\frac{S^{\prime}}{S}h^{\prime}_{k}+k^{2}{h}_{k}=0,
\end{equation}
where   prime means derivative with respect to the conformal time. Since the  polarization states are  same,   we here onwards denote  $h_{k}(\eta)$ without the polarization  index.  

Next, we  rescale the filed $h_{k}(\eta)$ by taking
$h_{k}(\eta)=f_{k}(\eta)/S(\eta)$, where the mode functions $f_{k}(\eta)$ obey the minimally coupled Klein-Gordon equation
\begin{equation}\label{zz}
f^{\prime \prime}_{k}+\Big(k^{2}-\frac{S^{\prime \prime}}{S} \Big)f_{k}=0.
\end{equation}
 The  general solution of the above equation  is a linear combination of  the Hankel function with a  generic power-law for the scale factor $S= \eta^{q}$  given by
\begin{equation}
f_k (\eta)= A_k \sqrt{ k \eta} H^{(1)} _{(q-\frac{1}{2})} (k \eta)+  B_k \sqrt{ k \eta} H^{(2)}_{(q-\frac{1}{2})} (k \eta).\end{equation}
For a given  model of the expansion of universe, consisting of a sequence of successive  scale factor with different $q$, we can obtain an exact solution  $f_k (\eta)$ by matching its value and  derivative at the joining points.

 The  approximate computation of the spectrum of  gravitational  waves  is usually performed in two limiting cases depending up on the waves that are within  or outside of the  barrier. For the  gravitational  waves  outside barrier ($k^{2}\gg S^{\prime \prime}/S$, short wave   approximation)  the corresponding   amplitude  decrease as $h_k \propto 1/S(\eta) $ while for the  waves inside the barrier ($k^{2} \ll S^{\prime \prime}/S$, long wave  approximation),  $h_k = C_k $ simply a constant. Thus these results can be used to estimate the spectrum for the present epoch of  universe.

The history of  overall expansion of the universe can be modeled as  following sequence
of successive epochs of power-law expansion.

The initial stage (inflationary)
\begin{equation}
S(\eta)=l_{0}|\eta |^{1+\beta},\;\;\;\;\;\;-\infty <\eta\leq \eta_{1},
\end{equation}
where $1+\beta <0$, $\eta<0$ and $l_{0}$ is a constant.

The z-stage
\begin{equation}
S(\eta)=S_{z}(\eta - \eta_{p})^{1+\beta_{s}},\;\;\;\;\;\;\eta_{1}<\eta\leq \eta_{s},
\end{equation}
where $\beta_{s}+1>0$. Towards the end of inflation, during the reheating, the equation of state of
energy in the universe can be quite complicated and is rather model-dependent \cite{qa}. Hence this
z-stage is introduced to allow a general reheating epoch, see for details \cite{3}.

The radiation-dominated stage
\begin{equation}
S(\eta)=S_{e}(\eta-\eta_{e}),\;\;\;\;\;\;\eta_{s}\leq \eta \leq \eta_{2},
\end{equation}

The matter-dominated stage
\begin{equation}
S(\eta)=S_{m}(\eta-\eta_{m})^{2},\;\;\;\;\;\;\eta_{2}\leq \eta \leq \eta_{E},
\end{equation}
where $\eta_{E}$ is the time when the dark energy density $\rho_{\Lambda}$ is equal to the matter energy density $\rho_{m}$. Before the discovery of  accelerating expansion of the universe, the current expansion is
used to take  as decelerating one  because of  the matter-dominated stage. Thus, following
the matter-dominated stage,  it reasonable to add an epoch of accelerating stage, which is probably driven
by either the cosmological constant, or the quintessence, or some other kind of condensate \cite{sa}. The value of  redshift $z_{E}$ at  $\eta_{E}$ is  $(1+z_{E})=S(\eta_{0})/S(\eta_{E})$, where $\eta_{0}$ is the present time.  Since $\rho_{\Lambda}$ is constant and $\rho_{m}(\eta) \propto S^{-3}(\eta)$, we get
\begin{equation}
\frac{\rho_{\Lambda}}{\rho_{m}(\eta_{E})}=\frac{\rho_{\Lambda}}{\rho_{m}(\eta_{0})(1+z_{E})^3}=1.
\end{equation}
If  the current value of $\Omega_{\Lambda}\sim0.7$ and $\Omega_{m}\sim0.3$, then it follows that
\begin{equation}
1+z_{E}=\Big(\frac{\Omega_{\Lambda}}{\Omega_{m}}\Big)^{1/3}\sim 1.33.
\end{equation}

The accelerating stage (up to the present)
\begin{equation}\label{1w}
S(\eta)=\ell_{0}|\eta- \eta_{a} |^{-1},\;\;\;\;\;\;\eta_{E}\leq \eta \leq\eta_{0}.
\end{equation}
This stage describes the accelerating expansion of the universe.  And  is a  new feature and hence  its  influence  on  the spectrum of relic gravitational waves is of interesting to study. It is be  noted  that the actual scale factor function $S(\eta)$ differs from equation (\ref{1w}), since
the matter component exists in the current universe. However, the dark energy is dominant,
therefore (\ref{1w}) is an approximation to the current expansion behaviour.

Given $S(\eta)$ for the various epochs, the derivative $S^{\prime}=dS/d\eta$ and  ratio $S^{\prime}/S$ follow
immediately. Except for  $\beta_{s}$ which is imposed upon as the model parameter, there are ten
constants in the  expressions of $S(\eta)$. By the continuity conditions of $S(\eta)$ and $S^{\prime}(\eta)$ at
 four given joining points $\eta_{1}, \eta_{s}, \eta_{2},$ and $\eta_{E}$, one can fix only eight constants. The remaining
two constants can be fixed by the overall normalization of $S$ and  the observed Hubble
constant as the expansion rate. Specifically, we put $|\eta_{0}-\eta_{a}|=1$ for the normalization of $S$, which fixes the  $\eta_{a}$, and the constant $\ell_{0}$ is fixed by the following calculation,
\begin{equation}
\frac{1}{H}\equiv \Big(\frac{S^{2}}{S^{\prime}}\Big)_{\eta_{0}}=\ell_{0}.
\end{equation}
where $\ell_{0}$ is  the Hubble radius at present. 

 In the expanding Friedmann-
Robertson-Walker spacetime the physical wavelength is related to the comoving wave
number  as
$\lambda \equiv \frac{2\pi S(\eta)}{k},$
and the wave number $k_{0}$ corresponding to the present Hubble radius is
$k_{0}=\frac{2\pi S(\eta_{0})}{\ell_{0}}=2\pi.$ And
there is another wave number
$k_{E}=\frac{2\pi S(\eta_{E})}{1/H}=\frac{k_{0}}{1+z_{E}},$
whose corresponding wavelength at the time $\eta_{E}$ is the Hubble radius $1/H$.

By matching $S$ and $S^{\prime}/S$ at the joint points, one gets
\begin{equation}\label{kk}
l_{0}=\ell_{0}b\zeta_{E}^{-(2+\beta)}\zeta_{2}^{\frac{\beta-1}{2}}\zeta_{s}^{\beta}\zeta_{1}^{\frac{\beta-\beta_{s}}{1-\beta_{s}}},
\end{equation}
where $b\equiv|1+\beta|^{-(2+\beta)}$, which is defined differently from  \cite{p}, $\zeta_{E}\equiv\frac{S(\eta_{0})}{S(\eta_{E})}$, $\zeta_{2}\equiv\frac{S(\eta_{E})}{S(\eta_{2})}$, $\zeta_{s}\equiv\frac{S(\eta_{2})}{S(\eta_{s})}$, and $\zeta_{1}\equiv\frac{S(\eta_{s})}{S(\eta_{1})}$. With these specifications, the functions $S(\eta)$ and $S^{\prime}(\eta)/S(\eta)$ are fully determined. In particular, $S^{\prime}(\eta)/S(\eta)$ rises up during the accelerating
stage, instead of decreasing as in the matter-dominated stage. This causes the modifications to the spectrum of relic gravitational waves.

\section{Gravitational waves spectrum in thermal vacuum state}
The power spectrum  of   gravitational waves is defined  as
\begin{equation}\label{pow}
\int_0 ^\infty h^2 (k,\eta) \frac{dk} {k} = \langle 0 | h^{ij}({\bf x},\eta) h_{ij}({\bf x},\eta) |0 \rangle,
\end{equation}
 Substituting equation  (\ref{1}) in (\ref{pow}) and taking the contribution from each polarization is same, we get
 \begin{equation}\label{pp}
 h(k,\eta)= \frac{4 l_{pl}}{\sqrt{\pi}} k \mid h(\eta) \mid.
 \end{equation} 
Thus once the mode function $h(\eta)$ is known, the spectrum $h(k,\eta)$ follows.

The spectrum at the present time $ h(k,\eta_0)$ can be obtained, provided the initial  spectrum is specified. The initial condition is taken to be the during the inflationary stage. Thus the initial amplitude of the spectrum is given by
\begin{equation}\label{bet}
h(k,\eta_i)= A{\left(\frac {k}{k_0}\right)}^{2+\beta},
\end{equation}
where $A=8\sqrt{ \pi} \frac{l_{pl}}{l_0} $ is a constant. The  power spectrum for the primordial perturbation of energy density  is $P(k)\propto {\mid  h(k,\eta_0)\mid}^2$ and in terms of  initial spectral index $n$, it is defined as $ P(k) \propto k^{n-1}$.
Thus the scale invariant spectral index  $n=1$ for the pure de Sitter expansion can be obtained   with the relation $n= 2 \beta +5 $  for $\beta $= - 2.

An effective approach to deals with the thermal  vacuum state is the thermo-field dynamics (TFD)\cite{34}. In this approach  a tilde space  is needed besides the usual
Hilbert space, and the direct product space is made up of the these two spaces. Every operator and state in the Hilbert space has the corresponding counter part in the tilde
space \cite{34}. Therefore a  thermal  vacuum state ($Tv$) can be  defined as
\begin{equation}\label{16}
|Tv\rangle ={\cal T }(\theta_{k})|0\; \tilde{0}\rangle,
\end{equation}
where 
 \begin{equation}\label{333}
{\cal T }(\theta_{k})=\mathrm{exp} [-\theta_{k} (a_{\bf {k}}\tilde{a}_{\bf {k}}-a_{\bf{k}}^{\dagger}\tilde{a}_{\bf {k}}^{\dagger})],
\end{equation}
is the thermal operator and $|0\; \tilde{0}\rangle$ is the two mode vacuum state at zero temperature. The  quantity $\theta_{k}$ is  related to   the average number of the thermal particle, $\bar{n}_{k}=\mathrm{sinh}^{2}\theta_{k}$. The $\bar{n}_{k}$  for given  temperature T is
provided   by the Bose-Einstein distribution 
$\bar{n}_{k}=[\mathrm{exp}( k /T)-1]^{-1}$, 
where $\omega_{k}$ is the resonance frequency of the field. The $a_{{\bf k}}$, $a_{{\bf k}}^{\dagger}$  and  $\tilde{a}_{{\bf  k}}$, $\tilde{a}_{{\bf k}}^{\dagger}$, are respectively the annihilation and creation operators in Hilbert and tilde space, satisfy the  usual commutation relations,  
$[a_{{\bf k}},a^{\dagger}_{{\bf k}^{\prime}}]=
[\tilde{a}_{{\bf k}},\tilde{a}^{\dagger}_{{\bf k}^{\prime}}]=\delta^{3}({\bf k}-{{\bf k}^{\prime}}) \; $. And all other commutation relations   of these operators are zero.
 By the appropriate action of the operator (\ref{333})  on $a_{{\bf k}}$ and $a_{{\bf k}}^{\dagger}$, we  get \cite{35}
\begin{eqnarray} \label{tt}
\nonumber {\cal T }^{\dagger}a_{\bf {k}}{\cal T }=a_{\bf {k}}\;\mathrm{cosh}\; \theta_{k} +\tilde{a}_{\bf {k}}^{\dagger}\; \mathrm{sinh} \; \theta_{k},\\
{\cal T }^{\dagger}a^{\dagger}_{\bf {k}}{\cal T }=a^{\dagger}_{\bf {k}}\; \mathrm{cosh}\; \theta_{k} +\tilde{a}_{\bf{k}}\;\mathrm{sinh}\; \theta_{k}.
\end{eqnarray} 
Hence  the  occupation number   in  thermal vacuum state
 can be written as
\begin{equation}\label{w}
\langle a^{\dagger}_{\bf {k}} a_{\bf {k} ^{\prime}}\rangle = \left( \frac{1}{e^{k/T} -1} \right)\delta^{3}(\bf {k}-\bf {k}^{\prime}).
\end{equation}
 Thus, using Eq.(\ref{1}) and Eqs.(\ref{16}-\ref{w}) in Eq.(\ref{pow}) the power spectrum in thermal vacuum state is obtained as
  \begin{equation}
 h^2 _T(k,\eta)=  \frac{16 l^2 _{pl} }{\pi} k^2 {\mid h(\eta)\mid}^2  \mathrm {coth}\Big[\frac{k}{2T}\Big],
   \end{equation}

Thus in comparison with Eq.(\ref{bet}), the spectrum  is expressed as
\begin{equation}
h(k,\eta_i)= A{\left(\frac {k}{k_0}\right)}^{2+\beta} \mathrm {{coth}^{1/2}\Big[\frac{k}{2T}\Big]}.
\end{equation}
   The  last  term becomes significant when the ratio $k/(2T)$ is less than unity. The  wave number $k$ and  temperature $T$ are comoving quantity  which are 
    related  to the physical parameters at the time of inflation, see for details \cite{25}. Thus it is expected an enhancement of the spectrum by a factor $\mathrm {coth}^{1/2}[{k}/{2T}] =\mathrm {coth}^{1/2}[{H S_i}/{2T_i}]$ .
   
   It is convenient to consider the amplitude of waves   in different  range of wave numbers \cite{15}.  Thus the amplitude  of the spectrum in thermal vacuum state  for  different ranges are given by
   
       (i) when $k\leq k_{E}$, the  corresponding wavelength is   greater the present Hubble radius. Thus the amplitude remain as the initial one and can be written as 

\begin{equation}\label{y}
h_{T}(k,\eta_{0})=A\Big(\frac{k}{k_{0}}\Big)^{2+\beta}\mathrm {{coth}^{1/2}\Big[\frac{k}{2T}\Big]},
\end{equation}

(ii)   the amplitude remains  approximately same as long as the wave inside the barrier   but begin to  decrease when it leaves the barrier by a factor $1/S(\eta)$, depending the value of scale factor at that time. This  process  continue until the barrier becomes higher than $k$ at a time  $\eta$ earlier than $\eta_0$, so the  amplitude has decreased by  the ratio of the scale factor at the time of  leaving the barrier $S_b$ to  its value at $\eta$, $S(\eta)$. This   is   in the   range  $k_{E}\leq k\leq k_{0}$.  
   
\begin{equation}\label{ke}
h_{T}(k ,\eta_{0})=A\Big(\frac{k}{k_{0}}\Big)^{\beta-1} \mathrm {{coth}^{1/2}\Big[\frac{k}{2T}\Big]}\frac{1}{(1+z_{E})^{3}}.
\end{equation}
Note  that this range is a new feature   on account of the  current acceleration of the universe  which is absent in the decelerating model as pointed out in \cite{15}.  The amplitude of     the waves  that left  the barrier at $S_b$ with waves numbers $k > k_0$  has  been decreased  up to the present time by a factor $ S_b / S( \eta _0)$. This affect the amplitude of the  present spectrum and is obtained as
\begin{equation}\label{sq}
h_{T}(k,\eta_{0})=A\Big(\frac{k}{k_{0}}\Big)^{2+\beta} \mathrm {{coth}^{1/2}\Big[\frac{k}{2T}\Big]} \frac{S_b}{S(\eta_0)}.
\end{equation}
This result  can be  used to obtain  the spectrum of the waves in the remaining range of wave numbers.

(iii) the  wave number that does not hit the barrier in the range $ k_{0}\leq k\leq k_{2} $ gives the amplitude as follows
\begin{equation}\label{l}
h_{T}(k,\eta_{0})=A\Big(\frac{k}{k_{0}}\Big)^{\beta} \mathrm {{coth}^{1/2}\Big[\frac{k}{2T}\Big]}\frac{1}{(1+z_{E})^{3}},
\end{equation}
 the spectrum  in this   interval  is differ from that of the matter dominated case by a the factor $\frac{1}{(1+z_{E})^{3}}$. The  wave lengths of the spectrum  in the range  are  long but smaller than the present Hubble radius.
 
 (iv) in  the range of wave number $ k _{2}\leq  k \leq k _{s}$ , the amplitude is
\begin{equation}\label{o}
h(k,\eta_{0})=A\Big(\frac{k}{k _{0}}\Big)^{1+\beta}\Big(\frac{k_{0}}{k_{2}}\Big) \frac{1}{(1+z_{E})^{3}}.
\end{equation}
 This is the interesting range on the observational point of view of Adv.LIGO, ET and LISA. Note that the temperature dependent factor in this  range  is negligible hence  the term  is dropped out  because of the low temperature nature of the  relic waves. 
 
 (v) for the wave number  range $k _{s}\leq k \leq k _{1}$  which is in  the high frequency  case and   gives the corresponding amplitude as
\begin{equation*}
h_{T}(k,\eta_{0})=A\Big(\frac{k}{k _{0}}\Big)^{1+\beta-\beta_{s}}\Big(\frac{k _{s}}{k _{0}}\Big)^{\beta_{s}}\Big(\frac{k _{0}}{k _{2}}\Big) \mathrm {{coth}^{1/2}\Big[\frac{k}{2T}\Big]}\frac{1}{(1+z_{E})^{3}}.
\end{equation*}
\begin{equation}\label{oo}
\end{equation}
 
 In the usual case the temperature dependent term can also be  neglected however the extra dimensional scenario  predicts  higher temperature for the  thermal gravitational waves, hence the term again becomes significant. Therefore the contribution from the thermal relic gravitational waves is  expected increase the amplitude of spectrum  particularly in the higher frequency range also.

\begin{figure}[t]
 {\includegraphics[scale=0.52]{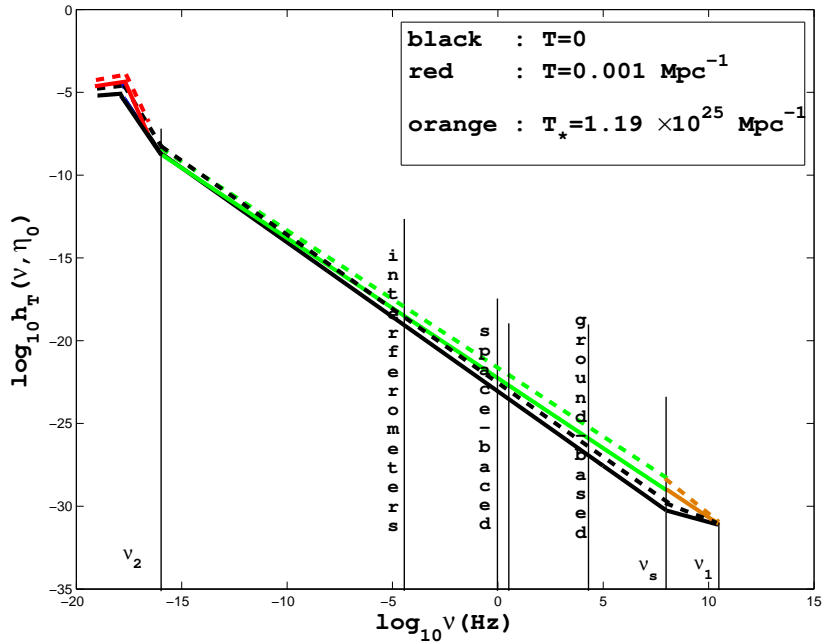}}
 \caption{ The  amplitude of  the gravitational waves  for the accelerated (solid lines) and decelerated (dashed lines) universe. }\label{ff1}
 \end{figure}

\begin{figure*}[t]
{\includegraphics[scale=0.52]{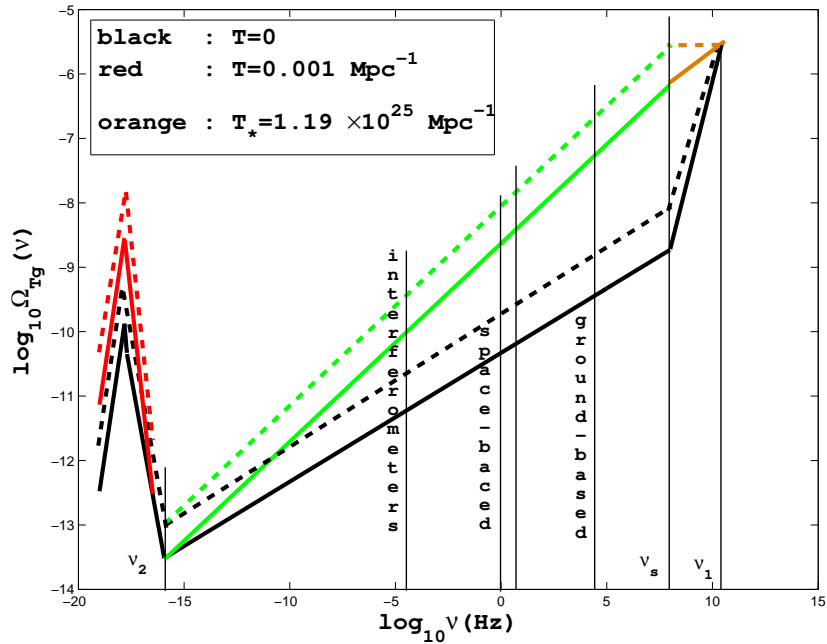}}
\caption{ The  spectral  energy density of the gravitational waves  for the accelerated (solid lines) and decelerated (dashed lines) universe.}\label{ff2}
\end{figure*}

 It is to be noted that in  (iv)   the  thermal contribution in $ k _{2}\leq  k \leq k _{s}$  range is  negligible   
due to  the temperature dependent  term.  Similarly  the thermal effect is  insignificant  in the range $k _{s}\leq k \leq k _{1}$ also. However
by taking into account  the extra dimensional effect,  the spectrum of  relic waves  is   peaked with a  temperature   $T_{*}$=1.19 $\times$ 10$^{25}$\;{Mpc}$^{-1}$ \cite{fry} (See, appendix  A for a brief discussion on $T_{*}$ from extra dimensional scenario.). Therefore it is expected  enhancement  for the amplitude of spectrum (orange lines, Figs.[\ref{ff1}] and [\ref{ff2}]) in the range  $ k _{s}\leq  k \leq k _{1}$ compared   to $T= 0$ case for the accelerated as well as  decelerated universe. But at the same time, ignoring  the  thermal contribution  to  the amplitude of  spectrum  in the range $ k _{2}\leq  k \leq k _{s}$ leads to a discontinuity  at $ k_{s}$, see Fig.[\ref{ff1}]. 
This   is evaded  by  fitting a new  line  in the range  $ k _{2}\leq  k \leq k _{s}$
for  the amplitude $h$  of Eq.(\ref{o})  as follows. 
 
 Let the amplitude of the wave in the range  $k _{0}\leq  k \leq k _{2}$ is given by  (\ref{l}) and can be rewritten as
\begin{equation}\label{1q}
h_{1T}(k,\eta_{0})=A\Big(\frac{k}{k_{0}}\Big)^{\beta} \mathrm {{coth}^{1/2}\Big[\frac{k}{2T}\Big]}\frac{1}{(1+z_{E})^{3}},
\end{equation}
and  the amplitude in the $k _{s}\leq  k \leq k _{1}$ is given by (\ref{oo}) also rewritten as

\begin{equation}
\label{3q}
h_{2T}(k,\eta_{0})=A\Big(\frac{k}{k _{0}}\Big)^{1+\beta-\beta_{s}}\Big(\frac{k _{s}}{k _{0}}\Big)^{\beta_{s}}\Big(\frac{k _{0}}{k _{2}}\Big) \mathrm {{coth}^{1/2}\Big[\frac{k}{2T_{*}}\Big]}\frac{1}{(1+z_{E})^{3}}.
\end{equation}
Thus the new slope for  Eq.(\ref{o}), in the range $ k _{2}\leq  k \leq k _{s}$, can be obtained 
by taking  $y\equiv \log_{10}(h)$ and $x\equiv \log_{10}(k)$, then 
\begin{equation}\label{poo}
\log_{10}(h)-\log_{10}(h)_{i}=\frac{\log_{10}(h)_{f}-\log_{10}(h)_{i}}{\log_{10}(k_{f})-\log_{10}(k_{i})}(\log_{10}(k)-\log_{10}(k_{i})),
\end{equation}
where the subscribes  $i$ and $f$  are respestively indicating  the first and last points of the straight line.
By putting $k_{i}\equiv k_{2}$ from Eq.(\ref{1q}) and $k_{f}\equiv k_{s}$ from Eq.(\ref{3q}) in   Eq.(\ref{poo}), we get \footnote{ here, $\mathrm {{coth}^{1/2}\Big[\frac{k_{2}}{2T}\Big]}=1. $}
\begin{equation}\label{p1}
h=(h_{1T})_{k_{2}}g(k),
\end{equation}
where
\begin{equation}
g(k)=\Big(  \frac{k}{k_{2}}\Big)^{\gamma},
\end{equation}
and
\begin{equation}\label{ss}
\gamma=\frac{\log_{10}(h_{2T})_{k_{s}}-\log_{10}(h_{1T})_{k_{2}}}{\log_{10}(k_{s})-\log_{10}(k_{2})}=\frac{\log_{10}\Big ( \Big(\frac{k_{s}}{k_{2}} \Big)^{1+\beta}  \mathrm {{coth}^{1/2}\Big[\frac{k_{s}}{2T_{*}}\Big]}\Big)}{\log_{10}\Big( \frac{k_{s}}{k_{2}}\Big)},
\end{equation}
is the slope of the line and thus  we find the amplitude, for convenience we  call it as `modified amplitude', given by
\begin{equation}\label{ppp}
h(k,\eta_{0})=A\Big (\frac{k_{2}}{k_{0}}\Big)^{\beta} \frac{1}{(1+z_{E})^{3}} \Big( \frac{k}{k_{2}} \Big)^{\gamma}.
\end{equation}
When $T_{*}$  becomes  zero    Eq.(\ref{ss}) leads to $\gamma=1+\beta$, and  hence   (\ref{o}) is recovered  from   (\ref{ppp})  in the range $k _{2}\leq  k \leq k _{s}$.

 The overall multiplication factor $A$  in all the spectra is determined in  absence of  the temperature dependent  term  with the CMB data of  WMAP  \cite{15}.  This is  based on the  assumption that the contribution from  gravitational waves  and  the density perturbations  are the same order of magnitude  or if  the CMB anisotropies at low multipole are induced by the gravitational waves,   therefore 
 it is possible to write  $\Delta T / T \simeq h(k,\eta_{0})$.  The observed CMB anisotropies \cite{u} at lower multipoles is $\Delta T / T \simeq0.37\times10^{-5}$ at $l\sim2$ which corresponds to the largest
scale anisotropies that have observed so far. Thus taking this to be the perturbations at the Hubble
radius  gives
\begin{equation}\label{k}
h(k_{0},\eta_{0})=A\frac{1}{(1+z_{E})^{3}}=0.37 \times 10^{-5}.
\end{equation}
However, there is a subtlety  in the interpretation of $ \Delta T / T$ at low multipoles, whose
corresponding scale is very large $ \sim \ell_{0}$. At present the Hubble radius is $\ell_{0}$, and the Hubble
diameter is $2\ell_{0}$. On the other hand, the smallest characteristic wave number is $k_{E}$, 
whose
corresponding physical wave length at present is $2\pi S (\eta_{0} )/k_{E}=\ell_{0}(1+z_{E})\simeq1.32 \ell_{0}$, which
is within the Hubble diameter $2 \ell_{0}$, and is theoretically observable. So, instead of Eq.(\ref{k}), if $ \Delta T / T \simeq 0.37 \times 10^{-5}$ at $l\sim2$ were taken as the amplitude of the spectrum at $\nu_{E}$, one would
have $h_T(k_{E},\eta_{0})=A/(1+z_{E})^{2+\beta}=0.37\times10^{-5}$, yielding a smaller $A$ than that in Eq. (\ref{k}) by a factor $(1+z_{}E)^{1-\beta}\sim2.3$ \cite{15}. The allowed values of  $\beta$ and $\beta_s$ are obtained and are  respectively give by $\beta=-1.9$, and $\beta_{s}=-0.552$ \cite{15}.

 \begin{figure}[t]
 {\includegraphics[scale=0.52]{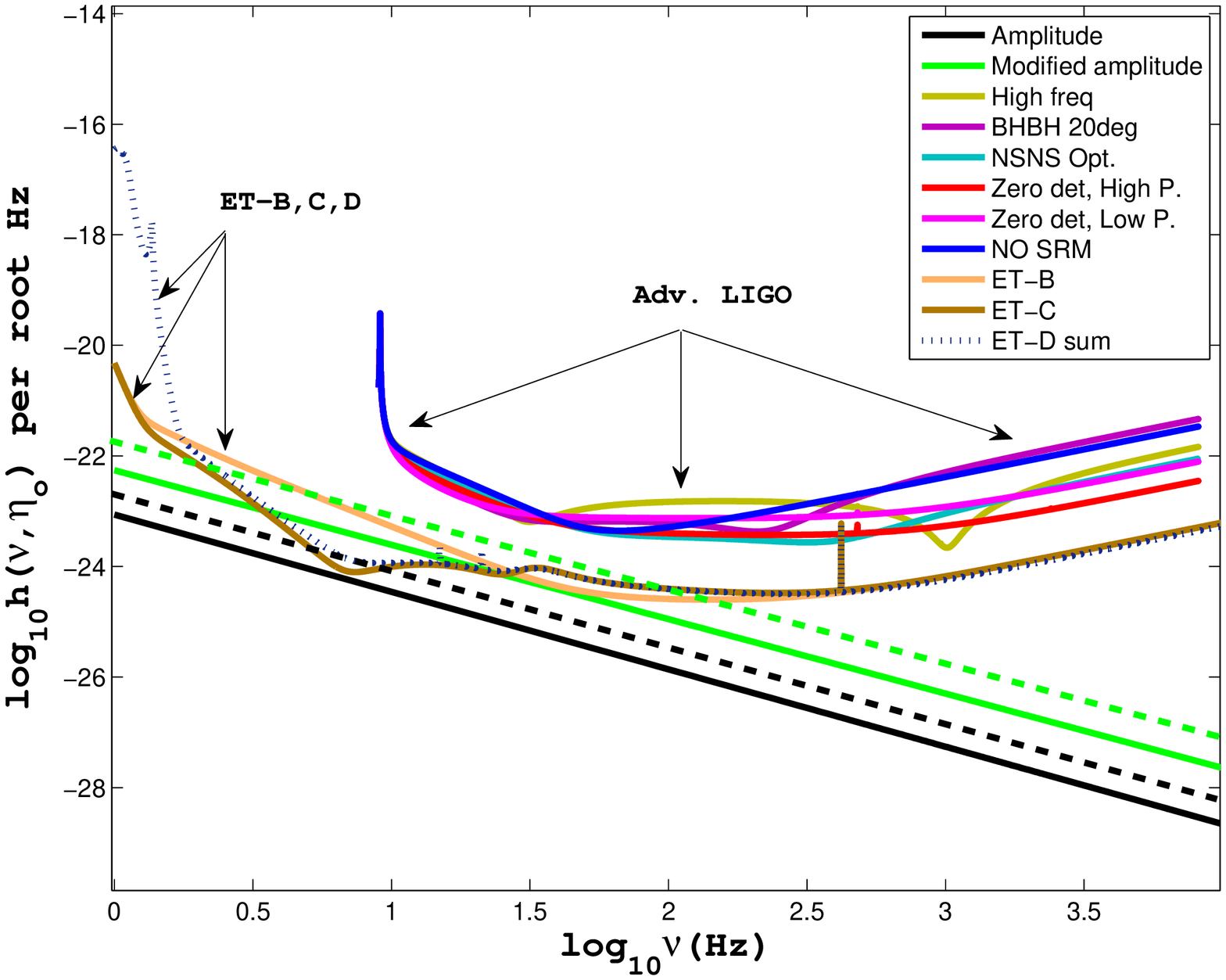}}
 \caption{  Comparison  of  the  modified  amplitude of the spectrum for the accelerated (solid black and green lines) and decelerated (dashed black and green lines) universe with  the  sensitivity curves of Adv.LIGO \cite{alg}   and ET \cite{et}.  }\label{f3}
\end{figure}    

 \begin{figure}[t]
 {\includegraphics[scale=0.52]{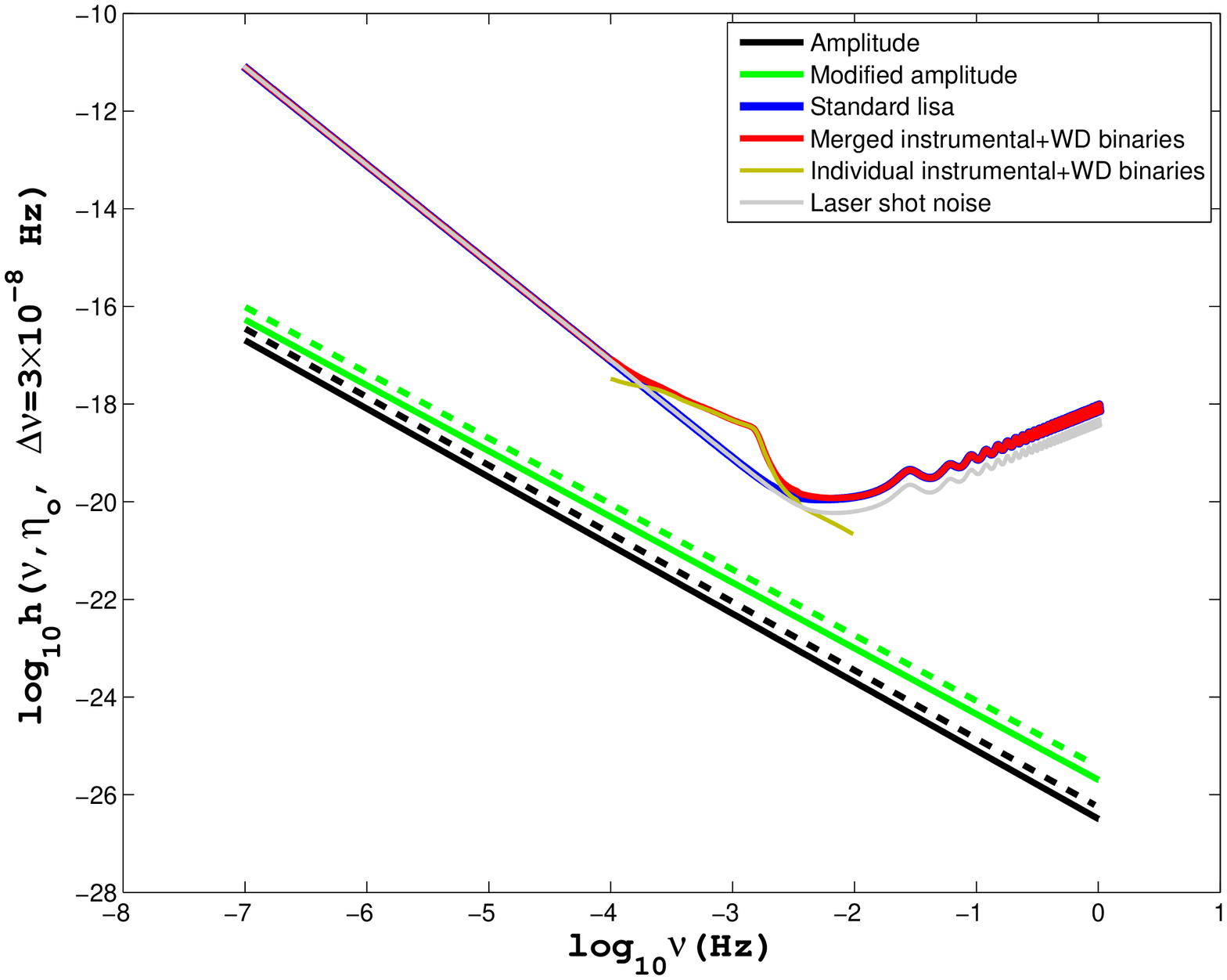}}
 \caption{Comparison  of  the  modified  amplitude of the spectrum for the accelerated (solid black and green lines) and decelerated (dashed black and green lines) universe with   the   LISA sensitivity curve. }\label{f4}
\end{figure}    
 
 Next, we obtain the spectrum in the thermal vacuum state with the following  parameters. By taking   $k=2\pi\nu$, $\nu_{E}=1.5\times10^{-18}$ Hz, $\nu_{0}=2\times10^{-18}$ Hz, $\nu_{2}=117\times10^{-18}$ Hz, $\nu_{s}=10^{8}$ Hz, $\nu_{1}=3\times10^{10}$ Hz ( the value of  $\nu_{1}$  is taken such a way that spectral energy density does not exceed the level of $10^{-6}$ , as required by the rate of  primordial nucleosynthesis). 
 The range of frequency is  chosen  in accordance with  generation of gravitational waves  that vary from early universe to  various astrophysical sources. And hence the range  is matching with  the interest   of CMB, Adv.LIGO, ET and LISA  operations for detection of  the gravitational waves.
The  spectrum is computed  in the thermal vacuum state  with the chosen values of the parameters  for  the  accelerated as well as decelerated model with  $T=0.001$Mpc$^{-1}$  in the low  range $k < k_{2}$. This temperature is    considered in the context of B mode of  CMB spectrum  in   thermal  state  \cite{25}. And  $T_{*}$  =1.19$\times$10$^{25}$Mpc$^{-1}$  \footnote{here,  $T_{*}$ = 0.905 K=  1.19$\times$10$^{25}$Mpc$^{-1}.$ } for the high  range $k_{s}\leqslant k\leqslant k_{1}$ which is from the   extra dimensional scenario \cite{fry}.  Since we use the natural unit, the wave number and temperature that appear in the temperature dependent term of the spectrum is computed  numerically in the Mpc$^{-1}$ unit.
The obtained spectra are normalized   of  the CMB  anisotropy spectrum of  WMAP data. 
 The  amplitude of the spectrum of  the thermal gravitational  waves is  enhanced  compared to its zero temperature case (vacuum case). It is  observed that the  spectrum for  $T=0.001$Mpc$^{-1}$ get maximum enhancement   $\sim 1.51$  times  than  the  vacuum case, at  $k=k_E$,
and   it is $\sim 20$ times for  $T_{*}$=1.19$\times$10$^{25}$Mpc$^{-1}$ at $k=k_s$.

 The  plots  for the  amplitude  of   spectrum $h_{T}(k,\eta_{0})$ versus the frequency $\nu$ for $\beta=-1.9$ and $\beta_{s}=-0.552$  are  given in Fig.[\ref{ff1}]. The amplitude of the  spectrum get enhanced  in the  frequency ranges, 10$^{-19}$ Hz$ \leq \nu <1.49 \times 10 ^{-17}$Hz, and $\nu_{s}\leq \nu\leq \nu_{1}$ ( the lower   value  of this range is  selected  such way that  thermal enhancement of the spectral density does not exceed the upper bound of nucleosynthesis rate.) due to the thermal effect of gravitational waves but  for the  range $ 1.49 \times 10 ^{-17} $Hz $\leq \nu < \nu_{s}$     there is a suppression because of  the $\coth^{1/2}[k/2T]$   term.  For  comparison, the amplitude of the spectra  are plotted for the decelerated and accelerated  universe, see  Fig.[\ref{ff1}].

 The new enhancement of the gravitational waves spectrum   due to the extra dimensional effect (the  modified amplitude, see  Fig.[\ref{ff1}], light green lines) can be compared with  the sensitivity of  Adv.LIGO, ET and LISA. An analytical expressions for the Adv.LIGO and ET  interferometers are  discussed in \cite{bs}. For  Adv.LIGO and ET cases, consider the root mean square amplitude per root Hz which equal to
 \begin{equation}
 \frac{h(\nu)}{\sqrt{\nu}}.
 \end{equation}
 The  comparison of the sensitivity  (10 Hz - 10$^4$ Hz) curve of the ground based interferometer  Adv.LIGO \cite{alg} with the  gravitational wave spectra  of  $\beta =-1.9$   for the accelerated and decelerated universe are given in Fig.[\ref{f3}].  Thus it shows that the Adv. LIGO  is unlikely to   detect  the enhancement   of the spectrum  from the  extra dimensional effect with its current stands but  be possible with  the sensitivity of  ET.

 Next, we compare the enhancement of the spectrum with the sensitivity  (10$^{-7}$ Hz - 10$^0$ Hz) of  space based detector LISA \cite{ss}. It is assumed that LISA  has one year observation time which corresponds to frequency bin $\Delta \nu$ = 3 $\times$ 10 $^{-8}$Hz ( one cycle/year) around each frequency.  Hence to make a comparison with the sensitivity curve,  a rescaling  of the spectrum $h(\nu)$ is required in Eq.(\ref{pp}) into the root mean square spectrum $h(\nu, \Delta \nu)$ in the band $\Delta \nu$, given by
 \begin{equation}
 h(\nu, \Delta \nu)=  h(\nu) \sqrt{\frac{\Delta \nu}{\nu}}.
   \end{equation}
 The plots of the LISA sensitivity  with the  modified amplitude of the spectrum are  given in Fig.[\ref{f4}] for the accelerated and decelerated universe. This show that the LISA is unlikely to detect the spectrum with the new enhancement feature of the gravitational waves.

The spectral energy density
parameter $\Omega_{g}(\nu)$ of gravitational waves is defined through the relation $\rho_{g}/\rho_{c}=\int\Omega_{g}(\nu)\frac{d\nu}{\nu}$, where $\rho_{g}$ is the energy density of the gravitational waves and $\rho_{c}$ is the critical energy density.
Thus
\begin{equation}\label{ka}
\Omega_{g}(\nu)=\frac{\pi^{2}}{3}h^2_{T}(k,\eta_{0})\Big(\frac{\nu}{\nu_{0}}\Big)^{2}.
\end{equation}
 Since the spacetime is assumed to be spatially flat 
$K=0$ with $\Omega=1$, the fraction density of relic gravitational waves must be less than unity, $\rho_{g}/ \rho_{c}<1$. After  
  normalization of  the spectrum by using Eq.(\ref{k}), we  integrate $\int\Omega_{g}(\nu)d\nu/ \nu$ from  $\nu_{*}=10^{-19}$ Hz up to  $\nu_{1}=3\times10^{10}$ Hz,  with $\beta=-1.9$ and $\beta_{s}=-0.552$. 
The integral is  evaluated   for the thermal case  and  zero temperature case, the  obtained results for the accelerated universe are  

(a) $\nu_{*}\leq\nu\leq\nu_{E}$,
\begin{eqnarray*}
\frac{\rho_{g}}{\rho_{c}}&=&5.8\times 10^{-11},\;\;\;\;T=0,\\
\frac{\rho_{g}}{\rho_{c}}&=&8.8\times 10^{-11},\;\;\;\;T=0.001\;{Mpc}^{-1},
\end{eqnarray*}

(b) $\nu_{E}\leq\nu\leq\nu_{H}$,
\begin{eqnarray*}
\frac{\rho_{g}}{\rho_{c}}&=&2.3\times 10^{-11},\;\;\;\;T=0,\\
\frac{\rho_{g}}{\rho_{c}}&=&3.5\times 10^{-11},\;\;\;\;T=0.001\;{Mpc}^{-1},
\end{eqnarray*}

(c) $\nu_{H}\leq\nu\leq\nu_{2}$,
\begin{eqnarray*}
\frac{\rho_{g}}{\rho_{c}}&=&2.4\times 10^{-11},\;\;\;\;T=0,\\
\frac{\rho_{g}}{\rho_{c}}&=&3.7\times 10^{-11},\;\;\;\;T=0.001\;{Mpc}^{-1},
\end{eqnarray*}

(d) $\nu_{2}\leq\nu\leq\nu_{s}$,
\begin{eqnarray*}
\frac{\rho_{g}}{\rho_{c}}&=&8.97\times 10^{-9},\;\;\;\;T=0,\\
\end{eqnarray*}

(e) $\nu_{s}\leq\nu\leq\nu_{1}$,
\begin{eqnarray*}
\frac{\rho_{g}}{\rho_{c}}&=&2.7\times 10^{-6},\;\;\;\;T=0. \\
\frac{\rho_{g}}{\rho_{c}}&=&6.67\times 10^{-6},\;\;\;\;T_{*}=1.19 \times10 ^{25}Mpc^{-1}.
\end{eqnarray*}

It is to be noted that in  (d)  the  thermal case are not shown  because  the  thermal contribution in this frequency range is  negligible   
due to  the temperature dependent  term. However
by taking into the extra dimensional effect, the  upper limit of temperature of the relic waves  is to be $T_{*}$=1.19$\times$10$^{25}$Mpc$^{-1}$. Thus it is expected an enhancement of the  spectral energy density in range  $\nu_{s}\leq\nu\leq\nu_{1}$ compared   to $T= 0$ case for the accelerated as well as  decelerated universe. But at the same time ignoring the thermal contribution  on the spectral density in the range  $\nu_{2}\leq\nu\leq\nu_{s}$ leads to a discontinuity  at $ \nu_{s}$, see Fig.[\ref{ff2}]. This problem  is solved  by  fitting a new  line  as discussed  in the context  of  estimation of the amplitude of the spectrum and  hence recomputed the spectral density in the range  $\nu_{2}\leq\nu\leq\nu_{s}$ which gives the new value $\frac{\rho_{g}}{\rho_{c}}$ = 8.21$\times 10^{-7}$. This 
 changes the slope  indicating enhancement of the spectral energy density of the gravitational waves in the range    $\nu_{2}\leq\nu\leq\nu_{s}$, green lines, Fig.[\ref{ff2}].
 
The enhancement spectral energy density $\Omega_{g}(\nu)$ in (d) can be  compared with  the estimated upper bound of various  studies and  are given  in Tab.[\ref{t1}].   Thus $\Omega^{(dec)}_{g}$ and $\Omega^{(acc)}_{g}(\nu)$   are less than the upper bound of  the estimated values  of   the respective frequency range.
\begin{table}
\caption{\label{t1}Comparison of  the estimated upper bound  of spectral energy density of  various studies  with   the present work.
Here $\Omega^{(dec)}_{g}$ and $\Omega^{(acc)}_{g} $ are   respectively the   spectral energy density  of the relic gravitational waves in  the decelerated and accelerated universe  of the present study and $\Omega^{(est)}_{g}$ is the  estimated upper bound of  various studies.
} 

\footnotesize\rm
\begin{tabular*}{\textwidth}{@{}l*{15}{@{\extracolsep{0pt plus12pt}}l}}
\br
Frequency($\nu$) Hz &$ \Omega^{(dec)}_{g}(\nu)$ & $\Omega^{(acc)}_{g}(\nu)$&  $\Omega^{(est)}_{g}(\nu)$\\
\mr
$10^{-9}-10^{-7} $ & $ 4.98 \times10^{-9}$ & $  1.03 \times10^{-9}$ & $2 \times 10^{-8} \;\;\;\cite{z}$ \\
$69-156$ &  $  34.84 \times10^{-8}$ & $  7.2 \times10^{-8}$ & $8.4\times 10^{-4}\cite{zz}$ \\
$41.5-169.25$  & $  4.93 \times10^{-7}$ & $ 1.02 \times10^{-7}$ & $6.9\times 10^{-6}
 \cite{zzz}$\\
\br
\end{tabular*}
\end{table}

Further  see that  the  contribution  to  $\rho_{g}/\rho_{c}$  from  the low frequency range  is $ {\cal{O}} (10^{-11}-10^{-10})$ while  from the higher frequency range it is $ {\cal{O}}(10^{-6})$.  Since the order of  contribution  to   the total $\rho_{g}/\rho_{c}$ from the lower frequency side is   very small  in  contrast with higher frequency side,  we  get  for the accelerated universe as
 \begin{equation}
 \rho_{g}/\rho_{c}\simeq 6.67\times10^{-6}   \; \; \;  \nu_{*}\leq\nu\leq\nu_{1},
\end{equation}
and is  the same order as that of the zero temperature case.
 However    $ \rho_{g}/\rho_{c}$  of the gravitational waves with $T\neq 0$ is higher  than  the zero temperature  case at lower frequency range $\nu_{*} \leq\nu\leq\nu_{2}$. Therefore  it is expected an enhancement for   the spectral energy density  in the thermal  vacuum state  in the  frequency range  $\nu_{*} \leq\nu\leq\nu_{2}$ and actually  it   is the range  of interest on the observational point of view of the  relic gravitational waves.   The total   estimated value of $ \rho_{g}/\rho_{c}$  by including    the thermal relic   gravitational waves in the very high frequency   does not alter the  upper bound of the nucleosynthesis rate. Thus the relic thermal  gravitons  with very high frequency range  are not ruled out and is   testable with  the upcoming  data of various missions  for detecting gravitational waves.
 \section{Discussion and conclusion}
 
Gravitational waves are  one of the classical predictions of  Einstein's general theory of relativity. The gravitational waves are generated  during the very early  evolution stages  of the universe as well as from the various astrophysical objects.  Therefore  frequency of the waves are  varying  very widely.  There are many on going  experiments  to detect these waves and the  interested range of frequency is from 10$^{-19}$ Hz to 10$^{10}$Hz.  The existence of gravitational waves with  very high  frequency range is  not  favoring by the energy scale of the conventional  inflationary scenario. However the  very high frequency range gravitational waves are interesting   candidates  in the models with extra dimensions.  The extra dimensional theories predict the existence of thermal gravitons with black body type spectrum.  These relic thermal gravitational waves can also add to the spectrum of the waves thus the corresponding amplitude  also  get enhanced. 
The nature of spectrum of the waves   to be observed today is   dependents on the  evolution history of the universe. Before the result of SN Ia observations,  the current evolution of the universe  is used to consider as  matter dominated  with decelerated expansion.  But, according  to the $\Lambda$CDM concordance model the present universe is supposed to be driven by dark energy  resulting an accelerated phase. If this  is the case then the spectral property of the waves  to be studied by taking into account of the current acceleration of the universe.    In the present work we  mainly considered the  very high frequency range  (The low frequency range thermal gravitational case  is carried out   by us without including the  very high frequency thermal waves  that comes from extra dimensional scenario and the   work is under  preparation. The enhancement of the lower frequency range is shown with red lines, see Figs.[\ref{ff1},\ref{ff2}]) of relic gravitational waves in the thermal vacuum state and obtained the spectrum for the accelerated as well as decelerated models.  The obtained spectra are normalized with the WMAP data. It is observed that  the inclusion of the very high relic thermal gravitational waves  leads to a discontinuity  in  the  amplitude of the spectrum at  $\nu_s$ (see Fig.[\ref{ff1}]). This is  
due to the fact that the temperature dependent term is insignificant in the  higher  frequency side of  the  range $\nu_2 \leq \nu \leq \nu_s$.  To evade this problem  a new  equation of  line  is derived and thus the amplitude get  enhanced in the range  $\nu_2  \leq \nu \leq \nu_s$. This is the new feature of the spectrum and we designates  it as the `modified amplitude' of the spectrum.
 The  modified amplitude of the spectrum can be compared with the sensitivity of  the Adv.LIGO, ET and LISA missions. The comparison of the Adv.LIGO  sensitivity shows that the modified amplitude   is unlikely to   detect    with its current stands of  LISA or   the improved sensitivity of Adv.LIGO. Where  as the  proposed sensitivity of  the ET  is promising to verify the modified amplitude  with its upcoming mission data.

The spectral energy density of the gravitational waves is  estimated   in thermal vacuum state for  the accelerated and decelerated universe.  It is observed that the spectral energy density get enhanced   in  the lower frequency range  $ {\cal{O}} (10^{-11}-10^{-10})$ and   from the higher frequency range it is $ {\cal{O}}(10^{-6})$. A comparison of  the estimated upper bound  of spectral energy density of  various studies  with   the present work is done. It shows that
 $\Omega^{(dec)}_{g}$ and $\Omega^{(acc)}_{g} $   are less than  the  estimated upper bound of  various studies.  The total   estimated value of $ \rho_{g}/\rho_{c}$  by including    the   very high frequency  thermal relic gravitational waves  does not alter the  upper bound of the nucleosynthesis rate. Thus the relic thermal  gravitons  with very high frequency range  are not ruled out and is   testable with  the upcoming  data of various missions  for detecting gravitational waves.

\section*{Appendix A}
 \section*{Extra dimensional Scenario and Thermal Gravitons} 
 
 Cosmology with extra dimensions have been motivated since Kaluza  and Klein (KK) showed that classical electromagnetism and general relativity could be combined in a five-dimensional framework. The modern scenarios involving extra dimensions are being explored in particle physics, with most models possessing either a large volume  or a large curvature. Although there exist  different  models of extra dimensions, there are some general features and signals common to all of them.

 In presence of  $D$ extra spatial dimensions, the 3+D+1- dimensional  action for gravity  for can be written as 
 \begin{equation}
 S = \int d^4 x \left [ \int d^D y \sqrt{- g_{D}} \frac{R_{D}}{16 \pi G_{D}}  + \sqrt{-g} L_{m} \right],
 \end{equation}
 where 
 \begin{equation}
 G_D = G_N \frac{m_{pl} ^2}{m_D ^{2+D}},
 \end{equation}
 and $g$ is the four dimensional metric, $G_N$  is  Newton's constant, $g_D, G_N$ and $R_D$ denote the higher dimensional counter parts of the metric, Newton's constant, and the Ricci scalar, respectively. $m_D $ is the fundamental scale of the extra dimensional theory. 
 
    Since the gravitational interactions are not strong enough to produce a thermal gravitons at temperatures below the Planck scale ($m_{pl}  \sim1.22 \times 10^{19}$ GeV), the standard  inflationary cosmology  predicted the existence the cosmic gravitational waves background  which are non-thermal in nature. However
 if the universe contains extra dimensions that can generate the thermal gravitational waves and its shape and amplitude of the CGWB may change significantly.  This can happen 
 when energies in the universe are higher than the fundamental scale $m_D$, the gravitational coupling strength increases significantly, as the gravitational field spreads out into the full spatial volume. Instead of freezing out at $\sim { \cal O}(m_{pl})$, as in 3+1 dimensions, gravitational interactions freeze-out at $\sim {\cal O}(m_D)$.  If the gravitational interactions become strong at an energy scale below the reheat temperature ($m_D  < T_{RH}$), gravitons get the opportunity to thermalize, creating a thermal CGWB. The qualitative result, the creation of a thermal CGWB if $m_D < T_{RH}$, is unchanged by the type of extra dimensions chosen \cite{fry}. 
 
 Thus, if extra dimensions do exist, and the fundamental scale of those dimensions is below the reheat temperature, a relic thermal CGWB ought to exist today. Compared to the relic thermal photon background (CMB), a thermal CGWB would have the same shape, statistics, and high degree of isotropy and homogeneity. The energy density ($\rho_g$) and fractional energy density ($\Omega_g$ )	of	a	thermal	CGWB	 are
 \begin{eqnarray}
 \rho_g = \frac{\pi^2}{15} \left( { \frac{3.91}{g_{\star}}} \right )^{4/3} T_{CMB}^4, \\
 \Omega_g = \frac{\rho_g}{\rho_c} \simeq 3.1 \times 10 ^{-4} (g_{\star})^{4/3},
 \end{eqnarray}
 where $\rho_c $ is the critical energy density today, $T_{CMB}$ is the present temperature of the CMB, and $g_{\star}$ is the number of relativistic degrees of freedom at the scale of $m_D$.  $g_{\star}$ is dependent on the particle content of the universe, i.e. whether (and at what scale) the universe is supersymmetric, has a KK tower, etc. Other quantities, such as the temperature (T), peak frequency ($\nu$), number density ($n$), and entropy density ($s$) of the thermal CGWB can be derived from the CMB if $g_{\star}$ is known, as 
 \begin{eqnarray}
 n_g = n_{CMB} \left( { \frac{3.91}{g_{\star}}} \right ),\,\, \, \, \, \,  s_g= s_{CMB}  \left( { \frac{3.91}{g_{\star}}} \right )\\
  T_g = T_{CMB} \left( { \frac{3.91}{g_{\star}}} \right )^{1/3},\,\, \, \, \, \,  \nu_g= \nu_{CMB}  \left( { \frac{3.91}{g_{\star}}} \right )^{1/3}.
    \end{eqnarray}
  These quantities are not dependent on the number of extra dimensions, as the large discrepancy in size between the three large spatial dimensions and the $D$ extra dimensions suppresses those corrections by at least a factor of  $ \sim 10^{ -29}$. If $m_D$ is just barely above the scale of the standard model, then $ g_{\star }$= 106.75. The thermal CGWB then has a temperature of 0.905 Kelvin, a peak frequency of 19 GHz \cite{fry}. 
  
  \section*{Acknowledgement}
Authors would like to thank S.Hild  for providing the ET sensitivity  data and  also Adv.LIGO and LISA web.


\section*{References}
\begin{thebibliography}{0}
\bibitem{gh} Grishchuk  L  1993 {\it Class. Quantum Grav.} \textbf{10} 2449; Grishchuk  L  2001, {\it Lect. notes Phys.} \textbf{562} 167
\bibitem{rex}Brax  P and  Bruck C  2003 {\it Classical Quantum Gravity} \textbf{20} R201 
\bibitem{ex}
Kaluza T {\it et al} 1921 {\it Math. Phys.} \textbf{1921} 966 ;
 Arkani  N-Hamed 1998 {\it Phys. Lett.B} \textbf{429}  263 ;
  N. Arkani-Hamed {\it et al} 1999 {\it Phys. Rev. D} \textbf{59}
 086004 ;  Randall L and Sundrum R 1999 {\it Phys. Rev. Lett.} \textbf{83}  3370 ; 
   Randall  L and  Sundrum R 1999 {\it Phys. Rev. Lett.} \textbf{83}  4690 
\bibitem{fry} Siegel  E R and Fry J N 2005 {\it Phys. Lett. B} \textbf{612} 122 ;  Pilo L,  Riotto A, and  Zaffaroni A 2004 {\it Phys. Rev. Lett.} \textbf{92} 201303 
\bibitem{24}  Abbott L F and  Wise M B,  1984 {\it Nucl. Phys. B} \textbf{244} 541 
\bibitem{25} Bhattacharya  K {\it et al} 2004 {\it Phys. Rev. Lett.} \textbf{97} 
251301; Zhao W {\it et al} 2009 {\it Phys. Lett. B} \textbf{680} 411 
\bibitem{29}  Gasperini M {\it et al} 1993 {\it Phys. Rev. D} \textbf{48} R439  
\bibitem{23}  Guth A H 1981 {\it Phys. Rev. D} \textbf{23} 347 ;  Linde A D
1982 {\it Phys. Lett. B} \textbf{108} 389 ; Albrecht  A and 
P J Steinhardt 1982 {\it Phys. Rev. Lett.} \textbf{48} 1220 
\bibitem{1}  Allen B 1988 {\it Phys. Rev. D} \textbf{37} 2078 
\bibitem{2} Sahni V  1990 {\it Phys. Rev. D} \textbf{42} 453 
\bibitem{3} Grishchuk  L  1997 {\it Class. Quantum Grav.} \textbf{14} 1445 
\bibitem{4}  Riszuelo A  and  Uzan J-P  2000 {\it Phys. Rev. D} \textbf{62} 083506 
\bibitem{5} Tashiro H, Chiba K  and Sasaki M 2004 {\it Class. Quantum Grav.} \textbf{21} 1761 
\bibitem{6}  Henriques A B 2004 {\it Class. Quantum Grav.} \textbf{21} 3057 
\bibitem{11} Riess A {\it et al}  1998 {\it Astron. J.} \textbf{116} 1009 
\bibitem{12} Perlmutter S {\it et al} 1999 {\it Astrophys. J.}\textbf{ 517} 565 
\bibitem{15} Zhang Y {\it et al} 2005 {\it Class. Quantum Grav.} \textbf{22} 1383–1394 
\bibitem{qa} Ford L H  1987 {\it Phys. Rev. D} \textbf{35} 2955
\bibitem{sa} Zhang Y  2002 {\it Gen. Rel. Grav.} \textbf{34} 2155  
\bibitem{p} Grishchuk L  {\it Testing Relativistic Gravity in Space} (Lecture Notes in
Physics vol 562)  Lammerzahl e d {\it et al} 2001 (Berlin: Springer ) p 164
\bibitem{34} Laplae L, Mancini F and Umezawa H (1974) {\it Phys. Rev. C} \textbf{10} 151 ; Takahashi Y, Umezawa H  1975 {\it Colloid Phenom.} \textbf{2} 55 ;
Umezawa H and Yamanaka Y 1988 {\it Adv. Phys.} \textbf{37} 531; Fearm H and Collett M J 1988 {\it J. Mod. Opt.} \textbf{35} 553 ; Chaturvedi  S {\it et al} 1990 {\it Phys. Rev. A} \textbf{ 41} 3969; 
Oz J-Vogt {\it et al} 1991 {\it J.Mod.Opt.} \textbf{38} 2339 
\bibitem{35}Lee C T 1990 {\it Phys. Rev. A} \textbf{42} 4193 ;
Xing-Lei Xu {\it et al} 2007 {\it Physica B.} \textbf{369}199 
\bibitem{u}  Spergel D N  {\it et al}  2003 {\it Astrophys. J. Suppl.} \textbf{148} 175 
\bibitem{alg} https://dcc.ligo.org/cgi-bin/DocB/showdocument
\bibitem{et} Einstein Telescope Web, http://www.et-gw.eu
\bibitem{bs} Mishra C K, Arun K G,  Iyer B R and Sathyaprakash B S 2010 {\it Phys. Rev. D}  \textbf{82} 064010
\bibitem{ss}http://www.srl.caltech.edu/˜ shane/sensitivty
\bibitem{z} Jenet  F A {\it et al} 2006 {\it The Astrophysical Journal} \textbf{653}1571-1576 
\bibitem{zz} Abbott B {\it et al} 2005 {\it Phys. Rev. Lett.} \textbf{95} 221101 
\bibitem{zzz}The Ligo Scientific Collaboration  2009 {\it nature} \textbf{460} 08278 

 \end{thebibliography}
\end{document}